\documentclass[twocolumn,superscriptaddress,longbibliography,prc]{revtex4-2}
\usepackage{times}
\usepackage[T1]{fontenc}

\usepackage{physics}
\usepackage{amsmath,amssymb,mathrsfs}
\usepackage[dvipdfm]{graphicx}
\usepackage{dcolumn}
\usepackage{bm}
\usepackage{morefloats}
\usepackage{multirow}
\usepackage{amssymb}
\usepackage{makecell}
\usepackage{amsmath}
\usepackage{xcolor}
\usepackage{longtable}
\usepackage{fix-cm}
\usepackage{verbatim}
\usepackage{float}
\usepackage[T1]{fontenc}
\usepackage[colorlinks,allcolors=blue]{hyperref}
\usepackage{booktabs}

\makeatletter
\def\NAT@def@citea{\def\@citea{\NAT@separator}}
\makeatother

\def\0\\{\nonumber\\}

\newcommand{\beq}{\begin{equation}}
\newcommand{\eeq}{\end{equation}}
\newcommand{\beqn}{\begin{eqnarray}}
\newcommand{\eeqn}{\end{eqnarray}}

\newcommand{\etal}{\textit{et al.}}
\newcommand{\eg}{\textit{e.g.}}
\newcommand{\ie}{\textit{i.e.}}
\newcommand{\cf}{\textit{cf.}}

\makeatletter
\newcommand\footnoteref[1]{\protected@xdef\@thefnmark{\ref{#1}}\@footnotemark}
\makeatother

\begin{document}


\title{
Fully general relativistic description of rapidly-rotating axially-symmetric neutron stars
for constraining nuclear matter equations of state
}

\author{Hyukjin Kwon}
\email{kwon.h.04c4@m.isct.ac.jp}
\affiliation{Department of Physics, School of Science, Institute of Science Tokyo, Tokyo 152-8550, Japan}

\author{Kazuyuki Sekizawa}
\email{sekizawa@phys.sci.isct.ac.jp}
\affiliation{Department of Physics, School of Science, Institute of Science Tokyo, Tokyo 152-8550, Japan}
\affiliation{Nuclear Physics Division, Center for Computational Sciences, University of Tsukuba, Ibaraki 305-8577, Japan}
\affiliation{RIKEN Nishina Center, Saitama 351-0198, Japan}

\date{\today}

\begin{abstract}
\edef\oldrightskip{\the\rightskip}
\begin{description}
\rightskip\oldrightskip\relax
\setlength{\parskip}{0pt} 
\item[Background]
Constraining the nuclear matter equation of state (EoS) from neutron star observations is one of the main subjects in nuclear physics today. The most of studies are based on the Tolman-Oppenheimer-Volkoff equation which describes a star in spherical hydrostatic equilibrium in general relativity. In general, however, neutron stars rotate rapidly and structure of neutron stars can be affected by rotation, especially in millisecond pulsars. To better constrain the nuclear EoS, it is important to describe neutron star structure taking into account the effects of rotation in a fully relativistic manner.

\item[Purpose]
In this study, we investigate the internal structure of neutron stars under the influence of rotation. We explore correlations between rotational effects and EoS parameters, based on fully general relativistic calculations of rapidly rotating neutron stars.

\item[Methods]
We consider the standard class of neutron stars with cores composed of neutrons, protons, electrons, and muons ($npe\mu$ matter), surrounded by the crust. For the EoS, we employ Skyrme-type energy density functionals for the $npe\mu$ matter and the conventional BPS EoS for the inner and outer crusts. To account for the rotational effects, we employ a method proposed by Komatsu, Eriguchi, and Hachisu (the KEH method), which provides stable solutions for axially-symmetric rotating equilibrium configurations.

\item[Results]
Using five different Skyrme-type EoS parameter sets, we find that the maximum angular frequency achievable by rotating neutron stars, as calculated via the KEH method, varies depending on the stiffness of the equation of state. We confirm that an increase in the rotating frequency leads to an overall increase in both the mass and radius along the $M$-$R$ curve, consistent with earlier studies. By performing calculations at three frequently referenced neutron stars, we further examine how the changes in mass and radius correlate with the nuclear matter properties at saturation density. Our results suggest that the often-quoted 716\,Hz rotational constraint may require a more conservative interpretation when accounting for realistic stellar deformation effects.

\item[Conclusions]
To place stringent constraints on the nuclear EoS based on observational data, it is sometimes essential to account for the effects of rotation in neutron star models. In particular, the influence of rotation becomes increasingly significant at higher spin frequencies and cannot be neglected in rapidly rotating systems with $\nu \gtrsim 400$\,Hz.

\end{description}
\end{abstract}
\maketitle

\section{Introduction}\label{Sec:Intro}

Neutron stars have been served as unique natural laboratories, spanning from sub-saturation density to ultra-high densities exceeding several times of the saturation density, where, in the latter case, even hyperon or quark degrees of freedom may appear. Inner structures of neutron stars are characterized by the equation of state (EoS), based on which one can obtain masses and radii of neutron stars and compare them with observational data. Consequently, neutron stars have been referred to as indispensable tools for constraining the structures and properties of dense matter under extreme conditions.

A large number of nuclear matter models have been proposed over the decades. For example, Dutra \etal~\cite{dutra2012,dutra2014} compiled 240 non-relativistic Skyrme-type and 263 relativistic mean-field (RMF) parametrizations. Similarly, Sun \etal~\cite{sun2018} surveyed 255 Skyrme-type and 270 RMF EoSs, as well as 13 non-relativistic Gogny-type ones. These models span a wide range of stiffness and symmetry energy behaviors, and their viability is frequently tested by comparing predicted neutron star properties against astrophysical observations.

The structure of a non-rotating (static), spherically symmetric neutron star can be computed by solving the Tolman–Oppenheimer–Volkoff (TOV) equation~\cite{oppenheimer1939}, a first-order differential equation derived from the Einstein field equations~\cite{Einstein1916}. While the TOV equation is widely used in the nuclear physics community due to its simplicity, it neglects rotational effects that are sometimes important for describing real neutron stars, especially rapidly rotating ones, like millisecond pulsars. In fact, many observed neutron stars are rapidly rotating. Notable examples include PSR\,J0030$+$0451~\cite{Miller2019} (205\,Hz), PSR\,J0740$+$6620~\cite{Wolff2021} (346\,Hz), and PSR\,J1748--2446ad~\cite{hessels2006} (716\,Hz), where the latter one is the fastest known pulsar to date. These millisecond pulsars provide empirical constraints on the maximum mass and stability of rotating configurations, particularly through their spin frequencies, which approach the Kepler (mass-shedding) limit.

To incorporate rotational effects into the description of neutron star structure, additional terms arising from centrifugal deformation and frame dragging must be considered. As the spin frequency increases, shape of a neutron star becomes gradually oblate, and the originally one-dimensional, spherically symmetric problem becomes a two-dimensional, axisymmetric one. These complications render the direct application of the TOV framework insufficient for modeling rapidly rotating neutron stars.

For slowly rotating configurations, the Hartle-Thorne perturbative approach~\cite{Hartle1967,Hartle1968} has been often adopted, in which the metric and fluid variables are expanded to the second order in angular velocity. While this formalism is analytically tractable and valid at low spin rates, a recent work by Kacskovics \etal~\cite{Kacskovics2022} has shown that significant discrepancies emerge between the Hartle-Thorne and a fully general relativistic models at spin frequencies exceeding about 600\,Hz. This underlines the need for fully relativistic numerical approaches for modeling rapidly rotating neutron stars.

For calculating rapidly rotating compact objects, several numerical approaches have been proposed to construct rotating stellar models, such as James’s method~\cite{James1964}, Stoeckly’s method~\cite{Stoeckly1965}, and the Eriguchi-Fukushima-Garabedian-Hachisu (EFGH) method~\cite{Eriguchi1985}. While these methods were useful in early studies, each of them has notable limitations. For example, James’s method could not construct models with polytropic indices $N>3$ or highly flattened configurations; Stoeckly’s method also failed to give solutions for the case of a high degree of flatness $T/|W|>0.25$, where $T$ and $W$ denote the total rotational energy and the gravitational energy, respectively; and the EFGH method, although capable of producing exotic shapes like a ``concave-hamburger'' or a ``dumbbell'' structure, was not applicable to configurations with discontinuous density distributions.

To address these limitations, Hachisu developed a Newtonian method in 1986, known as the Hachisu self-consistent field (HSCF) method~\cite{Hachisu1986}, which iteratively solves Poisson's equation to determine rotating equilibrium structures. This approach was later extended to general relativity by Komatsu, Eriguchi, and Hachisu in 1989, resulting in the KEH method~\cite{Komatsu1989}, which enables the computation of rapidly rotating stars in a fully general relativistic manner. A limitation of the original KEH formulation was its use of a non-compactified radial coordinate, requiring truncation of the computational domain at a large radius. To overcome this, Cook, Shapiro, and Teukolsky introduced a compactified radial coordinate $s \in (0,1)$ in 1992~\cite{Cook1992}, offering an efficient and accurate treatment of the entire spacetime, including the asymptotic region. This Cook-Shapiro-Teukolsky (CST) approach improves numerical convergence and resolution near the stellar surface.

While early applications of the KEH method employed polytropic EoSs for simplicity, Stergioulas and Friedman~\cite{Stergioulas1995} extended the framework to accommodate tabulated EoSs, facilitating more realistic modeling of dense matter. Their work culminated in the development of the open-source code, known as \texttt{RNS}, which combines the KEH and CST methods and supports arbitrary tabulated EoSs. The \texttt{RNS} code has since become a widely adopted tool in the neutron star community, enabling numerous studies on rapidly rotating configurations. Recent studies \cite{Qu2025,Konstantinou2024} have employed the \texttt{RNS} code to investigate a wide range of nuclear matter EoS under rapid rotation. For instance, in Ref.~\cite{Qu2025}, Qu \etal\ investigated the mass-radius relations of neutron stars under rotation using a relativistic \textit{ab initio} EoS based on the Bonn potential. Yeasin \etal~\cite{Yeasin2025} employed this code to investigate the properties of nuclear matter symmetry energy and the characteristics of neutron stars at the Keplerian frequency. In Ref~\cite{Rather2021}, Rather \etal\ studied the hadron-quark phase transition using RMF and bag model. In Ref.~\cite{Konstantinou2024}, Konstantinou extended the \texttt{RNS} code to a two-fluid model to explore the contribution of dark matter to rotational effects in neutron stars. 

Another well-known approach for computing rotating neutron stars is the Bonazzola–Gourgoulhon–Salgado–Marck (BGSM) method~\cite{Bonazzola1993}, which solves four coupled elliptic partial differential equations using a spectral method. This formulation is implemented in the open-source \texttt{Lorene} library. Despite its high numerical accuracy, the BGSM method involves considerable mathematical and numerical complexity. In contrast, the KEH method is more physically intuitive since it determines physical quantities sequentially, and its computational framework is more readily extensible. According to the work by Nozawa \etal~\cite{Nozawa1998,Stergioulas2003}, both approaches exhibit excellent agreement with each other under identical conditions.

However, \texttt{RNS} is fundamentally designed to compute macroscopic astrophysical quantities using tabulated inputs and does not directly provide access to microscopic properties such as internal density distributions or detailed composition profiles. Furthermore, plotting these internal structures or constructing mass–radius relations at a fixed rotation frequency, a task often required in nuclear-physics-oriented studies, requires a thorough understanding of the code and substantial modifications. Additionally, the code supports only the simplest rotational profile, namely rigid-body rotation, limiting its applicability in exploring more general differential rotation scenarios.

Aiming at investigating detailed properties of neutron star structures under rapid rotation and extending it to more general situations, like differential rotation, we have newly developed our own computational code based on the fully general relativistic framework of the KEH method, which can reproduce results from the \texttt{RNS} code. For the microphysical inputs, we adopt modern Skyrme-type nuclear energy density functionals (EDFs), which have been extensively calibrated to experimental data and are widely used in neutron star modeling. We select five representative parameter sets---BSk24~\cite{Goriely2013}, SkI4~\cite{REINHARD1995}, SkM*~\cite{BARTEL1982}, SkT5a~\cite{Duta}, and SLy4~\cite{Chabanat1998}---which are among the most commonly used in this field and span a range of nuclear matter and neutron star properties. We then perform direct numerical simulations across a range of spin frequencies, including those corresponding to observed millisecond pulsars, to examine rotational deformations, changes in the mass–radius relation, and their implications for nuclear EoS constraints.

The article is organized as follows. In Sec.~\ref{Sec:Methods}, we provide details of the theoretical models such as EoSs and the KEH method that are used in the present study. In Sec.~\ref{Sec:Details}, we summarize computational details and features of our code. In Sec.~\ref{Sec:Results}, we present numerical results of our calculations along with observational constraints and discuss the effects of rotation on neutron star properties. In Sec.~\ref{Sec:Conclusion}, a summary and a perspective are given.

\section{FORMULATION} \label{Sec:Methods}

\subsection{EoS parameters and laboratory constraints}

While a wide range of densities emerge inside a neutron star, only a limited density region around the saturation density is accessible in terrestrial experiments. It is therefore useful to characterize the nuclear matter EoS near the saturation density. To this end, it is customary to expand the energy per particle for a given nucleon number density $n$, $E(n)$, around the saturation density $n_0$. This expansion allows us to define a set of bulk nuclear matter parameters that encapsulate the response of the energy to density fluctuations and are essential for constraining EoSs. By expanding the energy per particle for the symmetric nuclear matter (SNM), \(E_\text{SNM}(n)\), in powers of the dimensionless variable, \((n-n_0) / 3n_0\), one obtains:
\begin{equation}
    \begin{aligned}
    E_{\text{SNM}}(n) &= E(n_0) + \left( \frac{\partial E(n)}{\partial n} \right)_{n_0} \left( \frac{n - n_0 }{3n_0}\right)  \\
    &\quad + \frac{1}{2} \left( \frac{\partial^2 E(n)}{\partial n^2} \right)_{n_0} \left( \frac{n - n_0 }{3n_0}\right)^2 \\
    &\quad + \frac{1}{6} \left( \frac{\partial^3 E(n)}{\partial n^3} \right)_{n_0} \left( \frac{n - n_0 }{3n_0}\right)^3 + \cdots.
    \end{aligned}
    \label{Eq:E_SNM_1}
\end{equation}
The coefficients of this expansion define key nuclear matter quantities at the saturation density, that is:
\begin{eqnarray}
    P_0 &=& \left( \frac{\partial E_{\text{SNM}}(n)}{\partial n} \right)_{n_0} = 0,
    \label{equation9}\\
    K_0 &=& 9n_0^2 \left( \frac{\partial^2 E_{\text{SNM}}(n)}{\partial n^2} \right)_{n_0},
    \label{equation10}\\
    Q_0 &=& 27n_0^3 \left( \frac{\partial^3 E_{\text{SNM}}(n)}{\partial n^3} \right)_{n_0}.
    \label{equation11}
\end{eqnarray}
Here, $P_0$ is the pressure at the saturation density $n_0$, which vanishes by definition, corresponding to the mechanical equilibrium. The coefficient $K_0$ represents the incompressibility of SNM, characterizing its stiffness, while $Q_0$ is the skewness parameter, describing the asymmetry of the EoS curvature around $n_0$. Using these definitions, the energy expansion \eqref{Eq:E_SNM_1} can be rewritten as
\begin{equation}
    \begin{aligned}
    E_{\text{SNM}}(n) &= E(n_0)
    + \frac{1}{2} K_0  \left( \frac{n - n_0 }{3n_0}\right)^2 \\
    &\quad + \frac{1}{6} 
    Q_0 \left( \frac{n - n_0 }{3n_0}\right)^3 + \cdots.
    \end{aligned}
    \label{Eq:E_SNM_2}
\end{equation}
These nuclear matter parameters can, in principle, be constrained by experimental observables and used to benchmark or constrain theoretical models.

Apart from the density derivative of EoS for SNM, an important quantity, the so-called symmetry energy, can be defined using two approaches. The first one is the difference between the energies of pure-neutron matter (PNM) and SNM; the second one involves the second derivative of the energy \(E(n, Y_p)\) with respect to the neutron excess \(1 - 2Y_p\) at \(Y_p = 1/2\), where $Y_p$ is the proton fraction. $Y_p= 1/2$ stands for SNM, \ie, equal numbers of protons and neutrons, which can be expressed as 
\begin{equation}
    S = \frac{1}{8} \left( \frac{\partial^2 E}{\partial Y_p^2} \right)_{Y_p=1/2}.
    \label{equation13}
\end{equation}
To characterize the behavior of the symmetry energy around the saturation density, one can apply a Taylor expansion similar to that for $E_\text{SNM}(n)$ \eqref{Eq:E_SNM_2}. Expanding \(S(n)\) around \(n_0\), we have:
\begin{equation}
    \begin{aligned}
        S(n) &= S_0 + L \left( \frac{n - n_0}{3n_0} \right)
        + \frac{1}{2} K_{\text{sym}} \left( \frac{n - n_0}{3n_0} \right)^2 \\
        &\quad + \frac{1}{6} Q_{\text{sym}} \left( \frac{n - n_0}{3n_0} \right)^3  \cdots,
    \end{aligned}
    \label{equation14}
\end{equation}
where $L$, $K_\text{sym}$, and $Q_\text{sym}$ are defined as follows:
\begin{eqnarray}
        L &=& 3n_0 \left( \frac{\partial S(n)}{\partial n} \right)_{n_0},
        \label{equation15}\\
        K_{\text{sym}} &=& 9n_0^2 \left( \frac{\partial^2 S(n)}{\partial n^2} \right)_{n_0},
        \label{equation16}\\
        Q_{\text{sym}} &=& 27n_0^3 \left( \frac{\partial^3 S(n)}{\partial n^3} \right)_{n_0}.
        \label{equation17}
\end{eqnarray}
The EoS parameters explained above are calculated for the five EoSs employed in this study and are given in Table~\ref{tab:nuclear_properties}.

\begin{table}[t]
\centering
\caption{Equation of state parameters at saturation density for five Skyrme parameter sets used in the present study. The saturation density $n_0$ is in the units of $\text{fm}^{-3}$, while all the other quantities are expressed in the units of MeV.}
\begin{tabular}{lccccccc}
\hline\hline
\textbf{Model} & \boldmath$n_0$ & \boldmath$K_0$  & \boldmath$Q_0$  & \boldmath$S_0$  & \boldmath$L$  & \boldmath$K_{\text{sym}}$  & \boldmath$Q_{\text{sym}}$  \\
\midrule
BSk24 & 0.157 & 244.95 & $-273.90$ & 29.99 & 46.52 & $-37.58$ & 708.11 \\
SkI4  & 0.160 & 247.95 & $-331.21$ & 29.50 & 60.39 & $-40.56$ & 351.16 \\
SkM*  & 0.160 & 216.61 & $-386.09$  & 30.03 & 45.78 & $-155.94$ & 330.47 \\
SkT5a  & 0.164 & 201.69 & $-436.81$  & 37.00 & 98.53 & $-24.97$ & 99.88 \\
SLy4  & 0.160 & 229.91 & $-363.11$ & 32.00 & 45.94 & $-119.73$ & 512.53 \\
\hline\hline
\end{tabular}
\label{tab:nuclear_properties}
\end{table}

$K_0$ represents the incompressibility of SNM, which is defined as the second derivative of the energy per particle with respect to density at saturation. It directly governs the pressure slope near the saturation density, $\left(\dd P/\dd n\right)_{n_0}$, and thus plays a crucial role in determining the stiffness of the EoS. Generally, a larger $K_0$ leads to a stiffer EoS around $n_0$. On the other hand, $L$ denotes the slope of the symmetry energy, and a larger $L$ implies a more rapid increase of $S(n)$ with density. This corresponds to a greater difference between neutron and proton chemical potentials at high densities. In $\beta$-equilibrated neutron star matter composed of $npe\mu$ components, an increase in $L$ tends to enhance the proton fraction by reducing the neutron excess $1-2Y_p$, which can soften the EoS compared to pure neutron matter (PNM). Indeed, the symmetry energy parameters, especially the slope of symmetry energy, $L$, are of particular importance in the astrophysical context, as they influence, \eg, neutron star radii, crust-core transition properties, and the tidal deformability observed in gravitational wave signals~\cite{Lattimer2023}.

The saturation density $n_0$ and the symmetry energy at saturation $S_0$ are relatively well constrained by laboratory experiments~\cite{Margueron2018}, typically around $0.16 \,\mathrm{fm}^{-3} $ and $ 30 \,\mathrm{MeV} $, respectively. In addition, both the incompressibility $K_0$ and the slope of the symmetry energy $L$ provide meaningful constraints, not yet so strict, but generally within $K_0 = 230 \pm 30 \,\mathrm{MeV}$~\cite{Shlomo2006,Piekarewicz2010} and $L = 60 \pm 30 \,\mathrm{MeV}$~\cite{Oertel2017,LI2013}. These quantities, in conjunction with neutron star observations, can provide valuable constraints on the nuclear EoS. We note, however, that parameters such as the skewness, $Q_0$, the incompressibility of the symmetry energy, $ K_\text{sym}$, and the skewness of the symmetry energy, $Q_\text{sym}$, remain loosely constrained and still span a wide range of possible values, making them less suitable for empirical calibration.

\subsection{Skyrme-type EoS for nuclear matter}

In this paper, we work with Skyrme-type EoSs for nuclear matter and here we briefly explain their details. A Skytme-type EoS can be derived, conventionally, within the Hatree-Fock theory with a Skyrme effective nucleon-nucleon interaction (Skyrme HF). The Skyrme forces, originally proposed by Skyrme in 1958~\cite{Skyrme1958}, describe nuclear interactions using a zero-range force represented by a delta function. The standard form of the Skyrme effective interaction is given by
\begin{equation}
    \begin{split}
    v(\boldsymbol{r}_1,\boldsymbol{r}_2) &= t_0 (1 + x_0 \hat{P}_\sigma)\delta(\boldsymbol{r}_1 - \boldsymbol{r}_2) \\
    &\quad + {1\over{2}}t_1 (1 + x_1 \hat{P}_\sigma) \{ \delta(\boldsymbol{r}_1 - \boldsymbol{r}_2) {\boldsymbol{k}}^2 + {\boldsymbol{k}}^{\prime2} \delta(\boldsymbol{r}_1 - \boldsymbol{r}_2) \} \\
    &\quad + t_2 (1+x_2 \hat{P}_{\sigma}){\boldsymbol{k}}' \boldsymbol{\cdot} \delta(\boldsymbol{r}_1 - \boldsymbol{r}_2) {\boldsymbol{k}} \\
    &\quad + {1\over{6}}t_3n^{\alpha}\left({\boldsymbol{r}_1+\boldsymbol{r}_2\over{2}}\right) (1+x_3 \hat{P}_{\sigma})\delta(\boldsymbol{r}_1 - \boldsymbol{r}_2) \\
    &\quad + iW_0({\boldsymbol{\sigma}}_1+{\boldsymbol{\sigma}}_2) \boldsymbol{\cdot} \{ {\boldsymbol{k}}' \boldsymbol{\times} \delta(\boldsymbol{r}_1 - \boldsymbol{r}_2) {\boldsymbol{k}} \}
    \label{equation1}
    \end{split}
 \end{equation}
with the operator $\boldsymbol{k}$ (and its conjugate $\boldsymbol{k}'$) acting on the right (left),
\begin{equation}
        \boldsymbol{k} = \frac{(\boldsymbol{\nabla}_1-\boldsymbol{\nabla}_2)}{2i}, \qquad \boldsymbol{k}' = -\frac{(\boldsymbol{\nabla}_1-\boldsymbol{\nabla}_2)}{2i}.
        \label{equation2}
\end{equation}
${\boldsymbol{\sigma}}$ denotes the Pauli spin matrices and the spin exchange operator is given by
\begin{equation}
    \hat{P}_{\sigma}={1\over{2}} (1+{\boldsymbol{\sigma}}_1 \cdot {\boldsymbol{\sigma}}_2).
    \label{equation3}
\end{equation}
The parameters $(t_0, t_1, t_2, t_3, x_0, x_1, x_2, x_3, \alpha, \text{and } W_0)$ are determined through a parameter fitting procedure based on selected properties of finite nuclei and nuclear matter, such as binding energies, radii, neutron skin thickness, saturation density, incompressibility, and many others.

In the Skyrme HF theory, the total energy of the system can be expressed as an expectation value of the Hamiltonian with the normalized Slater determinant~\cite{Vautherin1972},
\begin{equation}
    E_\text{Sky}=\langle\Psi | \hat{H}_\text{Sky} | {\Psi} \rangle=\int \mathcal{E}_\text{Sky}[n,\tau,\bm{j},\bm{s},\bm{T},J]\,\dd\bm{r},
    \label{equation4}
\end{equation}
where $\mathcal{E}_\text{Sky}$ is the energy density which depends on various local densities. For static situations the time-odd densities ($\bm{j}$, $\bm{s}$, and $\bm{T}$) vanish, and the spin-orbit density $J$ as well as gradient terms are absent when nuclear matter is uniform. By adopting the Thomas-Fermi approximation for the kinetic energy density, $\tau_q(\bm{r})=\frac{3}{5}(3\pi^2)^{2/3}n_q^{5/3}$, one can express the energy density as a functional of neutron and proton number densities alone, and it is often called the energy density functional (EDF).

In this research, we also adopt an extended form of the Skyrme interaction proposed by Chamel \etal~\cite{Chamel2009} for using BSk24, which includes density-dependent $t_4$ and $t_5$ terms, in order to eliminate the anomalous prediction of a ferromagnetic transition in neutron stars. The additional terms read
\begin{equation}
    \begin{split}
    &{1\over{2}}t_4 (1 + x_4 \hat{P}_\sigma)n^{\beta} \{ \delta(\boldsymbol{r}_1 - \boldsymbol{r}_2) {\boldsymbol{k}}^2 + {\boldsymbol{k}}^{\prime2} \delta(\boldsymbol{r}_1 - \boldsymbol{r}_2) \} \\
    &+ t_5 (1+x_5 \hat{P}_{\sigma}) n^{\gamma}{\boldsymbol{k}}'  \boldsymbol{\cdot} \delta(\boldsymbol{r}_1 - \boldsymbol{r}_2) {\boldsymbol{k}}.
    \end{split}
\end{equation}

For infinite asymmetric nuclear matter, the energy per particle is given by
\begin{equation}
    \begin{aligned}
        E(n,x) &=\frac{3\hbar}{10m} \left( \frac{3\pi^2}{2}  \right)^{2/3} n^{2/3} H_{5/3} \\
        &\quad + \frac{t_0}{4} n \left[ x_0+2- \left( x_0 + \frac{1}{2} \right) H_2 \right] \\
        &\quad + \frac{1}{24} t_3 n^{\alpha + 1} \left[ x_3 + 2 - \left( x_3 + \frac{1}{2} \right) H_2 \right] \\
        &\quad + \frac{3}{40} \left( \frac{3 \pi^2}{2}  \right)^{2/3} n^{5/3} (aH_{5/3} + bH_{8/3}) \\
        &\quad + \frac{3t_4}{40} \left( \frac{3 \pi^2}{2}  \right)^{2/3} n^{5/3+\beta} \\
        &\quad \times \left[ ( x_4 +2)H_{5/3} - \left(x_4 + \frac{1}{2} \right) H_{8/3} \right] \\ 
        &\quad + \frac{3t_5}{40} \left( \frac{3 \pi^2}{2}  \right)^{2/3} n^{5/3+\gamma} \\
        &\quad \times\left[ ( x_5 +2)H_{5/3} + \left(x_5 + \frac{1}{2} \right) H_{8/3} \right],
    \end{aligned}
    \label{equation5}
\end{equation}
where $m$ is the nucleon mass, and the parameters $a$ and $b$ are defined as
\begin{equation}
    \begin{aligned}
        a &= t_1 (x_1 + 2) + t_2 (x_2 + 2), \\
        b &= t_2 (x_2 + {1\over{2}}) - t_1 (x_1 + {1\over{2}}).
    \end{aligned}
    \label{equation66}
\end{equation}
The isospin dependence is encapsulated by the following function,
\begin{equation}
    \quad H_n(Y_p) = 2^{n-1}[Y_p^n + (1-Y_p)^n].
    \label{equation77}
\end{equation}
Notice that $H_n(Y_p=0)=2^{n-1}$ corresponds to PNM, while $H_n(Y_p=1/2)=1$ corresponds to SNM. From thermodynamic considerations, the pressure $P$ can be written as
\begin{equation}
    P=-\left( \frac{\partial E}{\partial V} \right)_{S,N} =-\left( \frac{\partial (N(\mathcal{E}/ n))}{\partial V} \right) = n^2 \frac{\partial (\mathcal{E}/ n)}{\partial n}.
    \label{equation99}
\end{equation}
The subscript $(S,N)$ implies that entropy and particle number are kept constant in differentiation. Thus, the pressure can be obtained analytically by differentiating the energy density with respect to the number density. The resulting analytic form of the pressure for the Skyrme EoS takes the following form:
\begin{equation}
    \begin{aligned}
        P(n,x) &=\frac{\hbar^2}{5m} \left( \frac{3\pi^2}{2}  \right)^{2/3} n^{5/3} H_{5/3} \\
        &\quad + \frac{t_0}{8} n^2 \left[ 2(x_0+2)- \left( 2x_0 + 1 \right) H_2 \right] \\
        &\quad + \frac{1}{48} t_3 (\alpha+1) n^{\alpha + 2} \left[ 2(x_3 + 2) - \left( 2x_3 + 1 \right) H_2 \right] \\
        &\quad + \frac{1}{8} \left( \frac{3 \pi^2}{2}  \right)^{2/3} n^{8/3} (aH_{5/3} + bH_{8/3}) \\
        &\quad + \frac{t_4}{40} \left( \frac{3 \pi^2}{2}  \right)^{2/3}(5+3\beta) n^{8/3+\beta} \\
        &\quad \times \left[ ( x_4 +2)H_{5/3} - \left(x_4 + \frac{1}{2} \right) H_{8/3} \right] \\
        &\quad + \frac{t_5}{40} \left( \frac{3 \pi^2}{2}  \right)^{2/3}(5+3\gamma) n^{8/3+\gamma} \\
        &\quad \times \left[ ( x_5 +2)H_{5/3} + \left(x_5 + \frac{1}{2} \right) H_{8/3} \right].
    \end{aligned}
    \label{equation6}
\end{equation}

The saturation density, denoted by $n_0$, is a fundamental property of SNM. It is defined as the density at which the pressure $P$ vanishes under symmetric conditions $(Y_p = 1/2)$, \ie,
\begin{equation}
    P(n_0, 1/2) = 0.
    \label{equation7}
\end{equation}
This condition corresponds to the equilibrium state of symmetric nuclear matter, where attractive and repulsive nuclear interactions are balanced, resulting in zero net pressure.

\subsection{Modeling of neutron star matter}

\begin{figure}[t]
    \centering
    \includegraphics[width=0.96\linewidth]{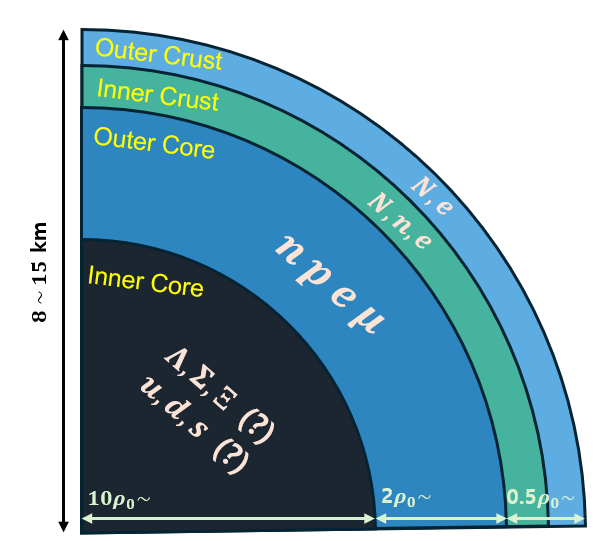}
    \caption{A schematic illustration of the cross section of a neutron star,
            showing its compositions and expected density ranges, where ``N''
            represents ``Nuclei'' forming Coulomb lattices in the outer and
            the inner crusts. In the present study, we do not consider the
            inner core, assuming that the entire core region is composed of
            $npe\mu$ nuclear matter.
            }
    \label{Figure1}
\end{figure}

To compute the structure of a neutron star, one must construct an EoS throughout the neutron star that reflects assumed internal compositions. Conventionally, the neutron star interior is divided into four regions: the outer crust, inner crust, outer core, and inner core. Figure~\ref{Figure1} provides a schematic illustration of the internal structure of a neutron star.

The outer crust consists of a Coulomb lattice of nuclei embedded in degenerated relativistic electron gas. As the density increases, going towards the inside of the neutron star, nuclei become neutron-rich through the electron capture process by protons, and eventually neutrons begin to drip out of nuclei. The latter region is called the inner crust, where Coulomb lattices coexist with gas of dripped superfluid neutrons. At even higher densities, matter transitions into a uniform phase of nuclear matter composed of neutrons, protons, electrons, and muons (commonly denoted as $npe\mu$ matter), which is referred to as the outer core.
Furthermore, in the deepest region of a neutron star, exotic degrees of freedom such as hyperons $(\Lambda, \Sigma, \Xi)$ or deconfined quarks $(u,d,s)$ may emerge, and such exotic phases are classified as the inner core.

In this study, we adopt the widely-used, classical Baym-Pethick-Sutherland (BPS) EoS~\cite{Baym1971} and Baym-Bethe-Pethick (BBP) EoS~\cite{Baym1971(2)} for the outer and inner crusts, respectively. For the core region, we assume a conventional $npe\mu$ composition of nuclear matter, disregarding the presence of exotic phases (the inner core), considering the fact that the presence of such exotic matter remains uncertain, offering an open question in nuclear astrophysics. The core is in a state of chemical equilibrium under weak interactions (beta equilibrium) and charge neutrality, satisfying the $\beta$-equilibrium condition,
\begin{subequations}
    \begin{align}
        \mu_n &= \mu_p + \mu_e, \label{equation19a} \\
        \mu_e &= \mu_\mu, \label{equation19b}
    \end{align}    
\end{subequations}
and the charge neutrality condition,
\begin{equation}
    n_p = n_e + n_\mu.
    \label{equation20}
\end{equation}

The total energy density of $npe\mu$ matter in the outer core is given by
\begin{equation}
    \varepsilon_{tot} = \varepsilon_\text{nucl} + \varepsilon_e + \varepsilon_\mu,
    \label{equation21}
\end{equation}
where $\varepsilon_\text{nucl}=m_nc^2n_n+m_pc^2n_p+\mathcal{E}_\text{Sky}$ denotes the nucleonic contribution to the energy density including the rest mass contributions, and $\varepsilon_l$ ($l=e$ or $\mu$) represents the leptonic contribution to the energy density, for which we use well-known formulas of relativistic Fermi gas. For a given baryon (nucleon) number density, the proton fraction is determined through the beta equilibrium and charge neutrality conditions, using the chemical potentials:
\begin{equation}
    \mu_q = \frac{\partial \varepsilon_\text{nucl}}{\partial n_q}\quad(q=n,p), \quad \mu_l = \varepsilon_l \quad(l=e,\mu).
\end{equation}
Using the particle fractions calculated under the $\beta$-equilibrium condition, we obtain the pressure and energy density as functions of baryon number density.

\subsection{Modeling of neutron star structure}

\subsubsection{Static case}

To calculate the structure of a neutron star in hydrostatic equilibrium within general relativity, we begin with the Einstein field equations, which relate the curvature of spacetime to the energy-momentum content of the star: 
\begin{equation}
    G_{\mu\nu} = R_{\mu\nu} - \frac{1}{2} g_{\mu\nu} R = \frac{8\pi G}{c^4} T_{\mu\nu},
    \label{equation23}
\end{equation}
where $G_{\mu\nu}$ is the Einstein tensor, $R_{\mu\nu}$ is the Ricci tensor, $g_{\mu\nu}$ is the metric tensor, $T_{\mu\nu}$ is the energy momentum tensor, and $G$ is the gravitational constant. We adopt the metric signature $(-,+,+,+)$ and a $c=G=1$ unit for general relativistic formulation given below. The matter inside the star is typically modeled as a perfect fluid, whose energy-momentum tensor is given by
\begin{equation}
    T^{\mu \nu} = (\varepsilon + P)u^{\mu}u^{\nu} + Pg^{\mu \nu},
    \label{equation24}
\end{equation}
where $\varepsilon$ is the total energy density, $P$ is the pressure, and $u^\mu$ is the fluid four-velocity.

Assuming spherical symmetry and static (\ie, non-rotating) equilibrium, the metric of the spacetime can be written as
\begin{equation}
    \dd s^2=-e^{2\nu(r)}\dd t^2+e^{2\lambda(r)}\dd r^2+r^2(\dd \theta^2+\sin^2\theta\dd\phi^2),
    \label{equation25}
\end{equation}
where $\nu(r)$ and $\lambda(r)$ represent metric functions. With this metric, and using the conservation row of the energy-momentum tensor, $\nabla_\mu T^{\mu\nu}=0$, one can derive the TOV equation:
\begin{eqnarray}
    \frac{\dd P}{\dd r} &=& - \frac{(\varepsilon+P)(M+4\pi r^3 P)}{r(r-2M)},
    \label{equation3.39}\\
    \frac{\dd M}{\dd r} &=& 4\pi r^2 \varepsilon.
    \label{equation3.38}
\end{eqnarray}

The TOV equation describes the structure of a spherically symmetric neutron star in equilibrium. However, for rapidly rotating neutron stars, such as millisecond pulsars, the centrifugal force significantly deforms the star and induces frame dragging. Thus, the system becomes axisymmetric, requiring a more general treatment.

\subsubsection{Rotating case}

To handle such rotating configurations, it is convenient to adopt a quasi-isotropic coordinate system. In this system, the line element takes the following form:
\begin{equation}
    \dd s^2 = -e^{\gamma+\varrho}\dd t^2 + e^{2\alpha}(\dd r^2 + r^2\dd\theta^2) + e^{\gamma-\varrho}r^2 \sin^2\theta(\dd\phi-\omega\dd t)^2,
    \label{equation4.14}
\end{equation}
where $\gamma(r,\theta)$, $\varrho(r,\theta)$, and $\alpha(r,\theta)$ are metric potentials, and $\omega(r,\theta)$ is the angular velocity of frame dragging. In the KEH method, which was proposed by Komatsu, Eriguchi, and Hachisu in 1989~\cite{Komatsu1989}, the Einstein equations are given by the following elliptic equations for the metric potentials:
\begin{subequations}
    \begin{align}
        &\nabla^2[\varrho e^{\gamma/2}] = S_{\varrho}(r,\mu), \\
        &\left( \nabla^2 + \frac{1}{r} \partial_r - \frac{\mu}{r^2} \partial_{\mu} \right) [\gamma e^{\gamma/2}] = S_{\gamma}(r,\mu), \\
        &\left( \nabla^2 + \frac{2}{r} \partial_r - \frac{2\mu}{r^2} \partial_{\mu} \right) [\omega e^{(\gamma-2\varrho)/2}] = S_{\omega}(r,\mu), \\
        & \frac{\partial \alpha}{\partial \mu} = S_\alpha(r,\mu),
    \end{align}
    \label{equation4.16}
\end{subequations}
where $\mu = \cos{\theta}$, and $\nabla^2$ denotes the flat-space Laplacian operator. The functions $S_\varrho, S_\gamma , S_\omega,$ and $S_\alpha$ are source terms, whose detailed expressions can be found in the original paper~\cite{Komatsu1989}. The proper velocity with respect to a zero angular momentum observer is given by~\cite{Butterworth1976}
\begin{equation}
    v = (\Omega-\omega)r \sin{\theta}e^{-\varrho},
\end{equation}
and the corresponding four-velocity reads
\begin{equation}
    u^\mu = \frac{e^{-(\varrho+\gamma)/2}}{\sqrt{1-v^2}} (1,0,0,\Omega).
\end{equation}
Here, $\Omega$ denotes the coordinate angular velocity.

The equation of hydrostatic equilibrium can be derived from the conservation of energy-momentum, $\nabla_\mu T^{\mu\nu}=0$:
\begin{equation}
    \dd P -(\varepsilon + P)\left[d\ln{u^t} + j(\Omega)\dd\Omega\right] = 0.
    \label{equation4.25}
\end{equation}
Here, the quantity $j(\Omega) = u^t u_\phi$ is the specific angular momentum, measured with respect to the proper time of the matter, and is assumed to be a function of the angular velocity $\Omega$ only. Under this assumption, Eq.~\eqref{equation4.25} can be integrated to yield the first integral of motion:
\begin{equation}
    \ln{H} + \frac{\varrho+\gamma}{2} + \frac{1}{2}\ln{(1-v^2)} + \int j(\Omega)\dd\Omega = C,
    \label{36}
\end{equation}
where $C$ is a constant and $H$ is the specific enthalpy defined as
\begin{equation}
    \ln{H} = \int \frac{\dd P}{\varepsilon + P}.
\end{equation}
The specific enthalpy $H$ is determined from the EoS.

In the KEH method, the functional form of the specific angular momentum is adopted based on the Newtonian limit:
\begin{equation}
    j(\Omega) = A^2 (\Omega_c - \Omega),
\end{equation}
where $A$ is a positive constant that controls the degree of differential rotation, and $\Omega_c$ denotes the angular velocity at the coordinate center. This assumption leads to the following relation:
\begin{equation}
    A^2(\Omega_c - \Omega)
    = \frac{(\Omega - \omega)\,r^2 \sin^2 \theta \,e^{-2\varrho}}{1 - (\Omega - \omega)^2\,r^2 \sin^2 \theta \,e^{-2\varrho}}.
    \label{39}
\end{equation}
To correctly describe differential rotation in a fully general relativistic framework, Eq.~\eqref{39} should be solved through a self-consistent iterative procedure. In the present study, however, we restrict ourselves to the limiting case of large $A$, which corresponds to rigid-body rotation, \ie, $\Omega = \Omega_c$. We will focus on exploring the relationship between rotational effects and properties of the nuclear EoS, leaving the extension to differential rotation for a future work.

Once the metric is determined by the self-consistent field method, the gravitational mass is calculated through the following relation:
\begin{equation}
    \begin{aligned}
        M &= 2\pi \iint \dd r\dd\theta r^2 \sin{\theta} e^{2\alpha+\gamma} \\
        &\quad \times \left[ \left( \frac{(\varepsilon+P)(1+v^2)}{(1-v^2)} + 2P \right) + 2r\sin{\theta} \omega e^{-\varrho} \frac{(\varepsilon+P)v}{(1-v^2)} \right].
    \end{aligned}
\end{equation}

To compute the Kepler frequencies, the solution of the geodesic equation yields two possible values for the orbital three-velocity~\cite{Bardeen1970,Cook1994}:
\begin{equation}
\tilde{v} =
\frac{
    e^{-\varrho} r^2 \omega_{,r}
    \pm
    \sqrt{
        e^{-2\varrho} r^4 \omega_{,r}^2
        + 2r(\gamma_{,r} + \varrho_{,r})
        + r^2\left(\gamma_{,r}^2 - \varrho_{,r}^2\right)
    }
}{
    2 + r(\gamma_{,r} - \varrho_{,r})
},
\end{equation}
where the comma notation denotes partial derivatives with respect to the radial coordinate $(\eg, \omega_{,r}\equiv\partial\omega/\partial r)$. The positive sign corresponds to corotating orbits, while the negative sign corresponds to counter-rotating orbits with respect to a zero-angular-momentum observer. The Keplerian angular frequencies (the mass-shedding limit), where matter can escape from the gravitational potential, derived in the general relativistic formalism are then defined by~\cite{Sharon1999}:
\begin{equation}
    \Omega_{K \pm} = \left[ \frac{ \tilde{v}_{\infty} }{2\pi r} \right]_{\pi / 2} =\left[ \frac{1}{2 \pi} \left( \tilde{v} \frac{e^\varrho}{r} + \omega \right) \right]_{\pi /2}.
    \label{Eq:Keplar_GR}
\end{equation}

\section{Computational details}\label{Sec:Details}

We have newly developed our own computational code from scratch to calculate structure of rotating neutron stars employing the KEH method with various Skyrme-type nuclear EoSs. The code basically follows the computational scheme described in the original paper \cite{Komatsu1989}. We have tested the code against available data, \eg, computed stellar structures using tabulated EoSs show excellent agreement with the results of Stergioulas and Friedman~\cite{Stergioulas1995}, and computed Kepler frequencies are consistent with those reported in Ref.~\cite{Sharon1999}. Since now we know the details of the formalism and the code, we can flexibly modify or extend the code to get deeper insight into physics behind.

The equilibrium condition, Eq.~\eqref{36}, is solved self-consistently. In the present work, we adopt the CST approach to address the computational truncation problem. The radial coordinate $r$ is mapped to the so-called compactified coordinate $s$ $(0 \leq s \leq 1)$, such that $r=\infty$ corresponds to $s=1$. A detailed description of this method can be found in Ref.~\cite{Cook1992}. In the code, we discretize the computational domain spanned by the compactified radial coordinate $s$ and the angular coordinate $\mu = \cos{\theta}$ into $257\times257$ grid points. The integrals of the Poisson-like field equations are evaluated using the Simpson's rule, in which the spectral expansion with Legendre polynomials up to $P_{32}(\mu)$ are used.

For the nuclear EoS, we employ five selected Skyrme EoSs, \ie, BSk24~\cite{Goriely2013}, SLy4~\cite{Chabanat1998}, SkM*~\cite{BARTEL1982}, SkT5a~\cite{Duta}, and SkI4~\cite{REINHARD1995}, in the present study. For each of the five parameter sets, we have confirmed that the properties of nuclear matter at saturation, such as the incompressibility, the skewness, and the symmetry energy slope, are consistent with the values reported in the original references. Prior to self-consistent KEH calculations, we construct tabulated EoS for $npe\mu$ matter with density intervals of $0.01\,\text{fm}^{-3}$. During the calculation, the tabulated EoS is interpolated using the logarithmic linear interpolation to ensure smooth and accurate evaluation of thermodynamic quantities.

\begin{table}[t]
\caption{Results of the KEH calculations for $R_\text{ratio}=R_\text{p}/R_\text{e}=0.8$ and central density $\rho_c = 1.5 \times 10^{15}$\,g/cm$^3$ ($n_c\simeq4.75n_0$) with SLy4 EoS, with five mesh resolutions, $129 \times 129$, $257 \times 257$, $513\times513$, $1025\times1025$, and $2049\times2049$.
}\vspace{2mm}
\begin{tabular}{lccc}
\hline\hline
\# of grid points & Mass ($M_\odot$) & Radius (km) & Angular Velocity (Hz) \\
\hline
$2049\times2049$ & 2.02776 & 12.278660 & 1127.40 \\
$1025\times1025$ & 2.02779 & 12.278641 & 1127.33 \\
 $513\times513$  & 2.02788 & 12.278580 & 1127.33 \\
 $257\times257$  & 2.02817 & 12.278202 & 1127.16 \\
 $129\times129$  & 2.02947 & 12.277360 & 1126.70 \\
\hline\hline
\end{tabular}\label{tab:resolutions}
\end{table}

In practice, the complexity of solving for the equilibrium configuration of rotating neutron stars is so high that some minor differences may arise depending on numerical conditions. For instance, even within the same KEH method, discrepancies can appear depending on the coordinate choices~\cite{Nozawa1998}. A change of the mesh resolution~\cite{Stergioulas1995} was shown to affect the results with less than 0.1\%. In addition, small uncertainty may arise from the interpolation scheme or other numerical details in the code. As an example, the equilibrium values, calculated for the same configuration with $R_\text{ratio}\equiv R_\text{p}/R_\text{e}=0.8$, where $R_\text{p}$ and $R_\text{e}$ denote the polar and equatorial radii of a deformed neutron star, respectively, and central density $\rho_c = 1.5\times10^{15}$\,g/cm$^3$ ($n_c\simeq 4.75n_0$) with SLy4 parameter sets, but with different mesh resolutions, $129 \times 129$, $257 \times 257$, $513\times513$, $1025\times1025$, and $2049\times2049$ are shown in Table~\ref{tab:resolutions}.
From the table, we find that the mass and radius agree up to the third decimal place except the $129\times129$ case. We thus fixed the mesh resolution to $257\times257$ and investigate the effects of rotation throughout the calculations in the following analyzes.

\begin{figure}[t]
    \centering
    \includegraphics[width=\linewidth]{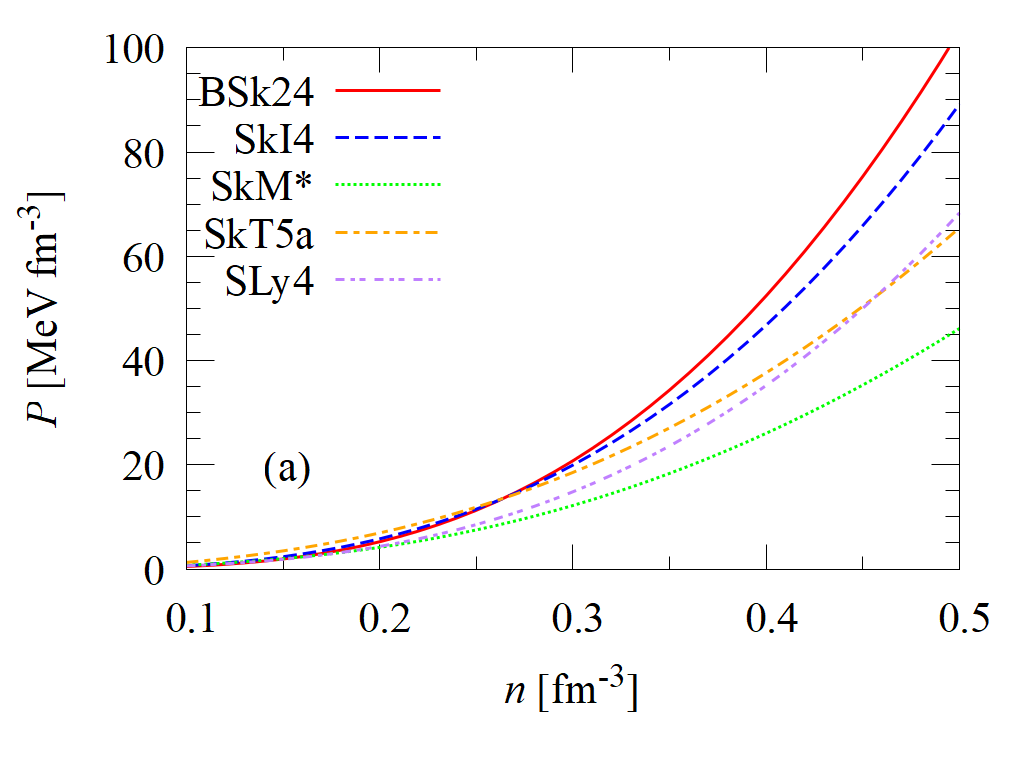}\vspace{-5mm}
    \includegraphics[width=\linewidth]{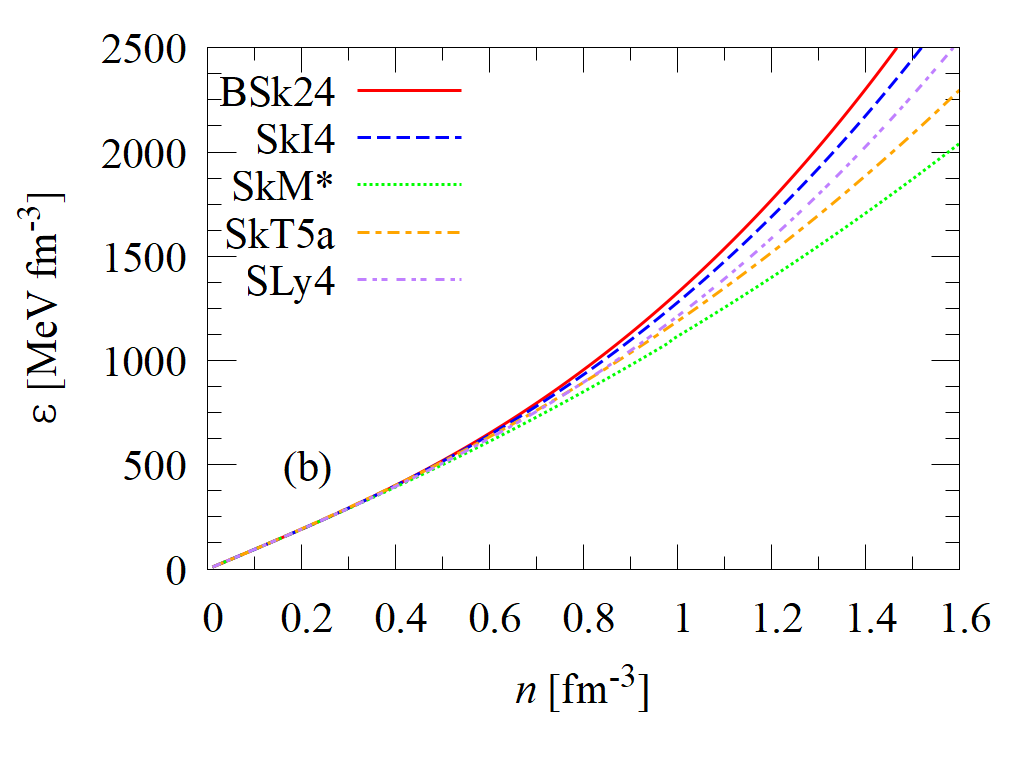}\vspace{-6mm}
    \caption{
    Equations of state calculated under the $\beta$-equilibrium condition using five Skyrme parameter sets. In panels (a) and (b), the pressure $P$ and the energy density $\varepsilon$ are shown, respectively, as a function of baryon number density $n$. Results are shown for five Skyrme parameter sets: BSk24 (solid line), SkI4 (dashed line), SkM* (dotted line), SkT5a (dash-dotted line), and SLy4 (two-dot-chain line).
    }
    \label{fig:EoS}
\end{figure}

\section{Results and Discussion}\label{Sec:Results}

\subsection{Nuclear equations of state}

It is helpful to have a careful look into the characteristics of the nuclear EoSs, before investigating the rotational effects on the neutron star structure. The resulting EoSs for the five Skyrme parameter sets examined are shown in Fig.~\ref{fig:EoS}. As can be seen from the figure, the five parameter sets span a wide range of EoS parameters showing different density dependence, especially for high density region. In the figure, the pressure and the energy density are shown in panels (a) and (b), respectively, as a function of baryon number density $n$. From the figure, we find that BSk24 yields the stiffest EoS, while SkM* produces the softest one, among the five outer core EoSs examined. We note that SkI4 has a higher incompressibility ($K_0\simeq248$\,MeV) than BSk24 ($K_0\simeq245$\,MeV), as shown in Table~\ref{tab:nuclear_properties}. One may thus naively expect that SkI4 would provide the stiffest EoS among the five examined. However, since SkI4 has a substantially larger $L$ value ($L\simeq60$\,MeV) as compared to BSk24 ($L\simeq47$\,MeV), the proton fraction gets higher under the $\beta$-equilibrium at higher densities, thereby reducing the pressure, resulting in a softer EoS than BSk24. Conversely, SkT5a exhibits a relatively small incompressibility ($K_0\simeq202$\,MeV), but due to its large symmetry energy ($S_0=37$\,MeV) and its slope ($L\simeq99$\,MeV), the reduction in neutron excess occurs rapidly at low densities. At high densities, the pressure is predominantly determined by the magnitude and slope of the symmetry energy, enabling SkT5a to maintain a stiffer EoS than SkM*, despite its lower $K_0$ value.

Based on these EoSs, we perform neutron star structure calculations to determine the internal configuration of rotating neutron stars, whose details are given in the subsequent sections.

\subsection{Effects of rotation on neutron star structure}

Next, let us investigate effects of rotation on neutron star structure and discuss correlations with the nuclear EoS parameters. The KEH method computes the equilibrium structure of a rotating neutron star by taking the central density and the axis ratio between the polar and equatorial radii as inputs, and solving for the metric potentials and the corresponding angular frequency. It means that the angular frequency is determined self-consistently based on the equilibrium configuration permitted by the given EoS.

\begin{figure}[t]
    \centering
    \includegraphics[width=\linewidth]{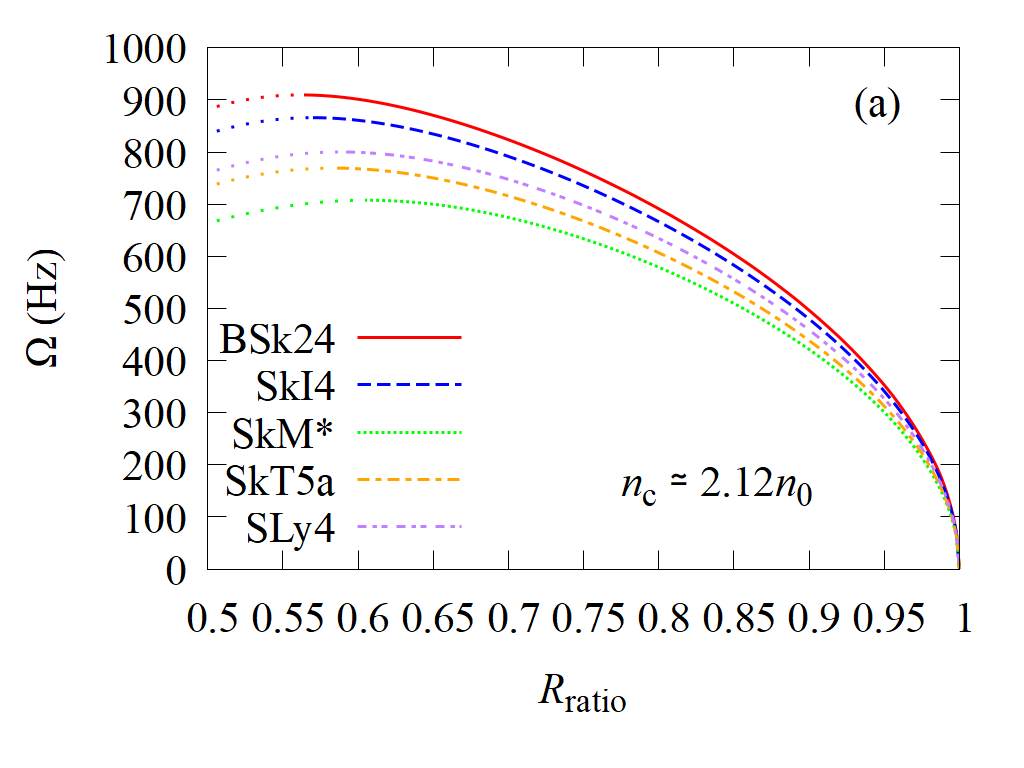}\vspace{-4mm}
    \includegraphics[width=\linewidth]{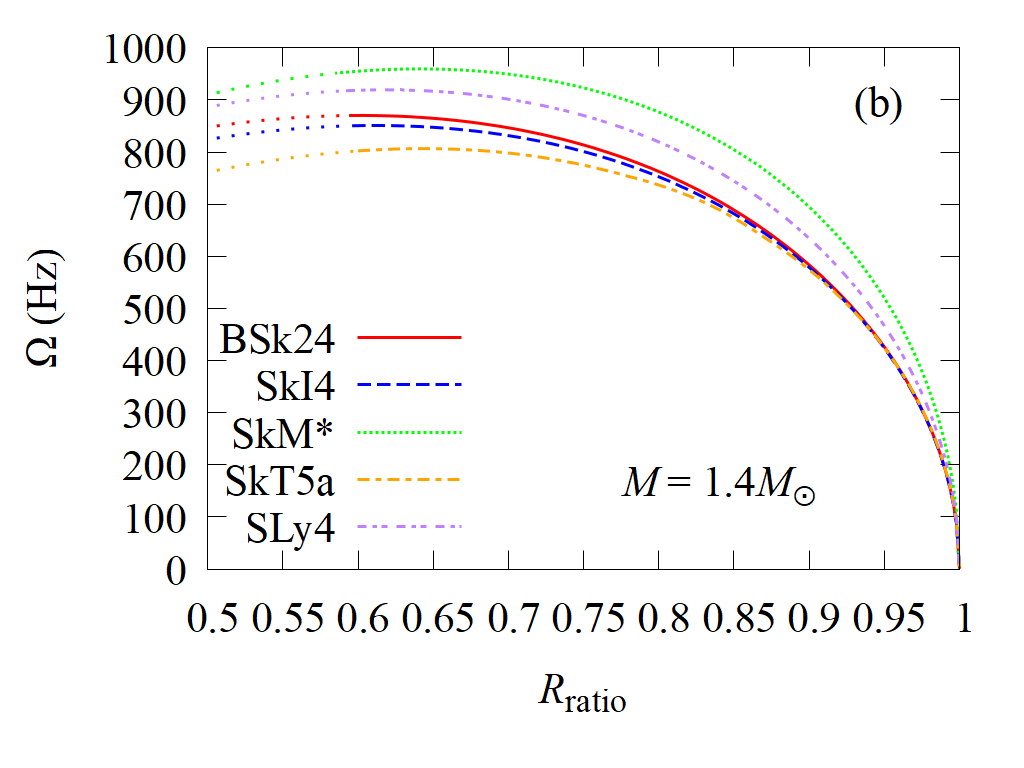}\vspace{-6mm}
    \caption{
    Angular frequency \( \Omega \) is shown as a function of the axis ratio \( R_{\text{ratio}} = R_\text{p}/R_\text{e} \), where \( R_\text{p} \) and \( R_\text{e} \) are the polar and equatorial radii, respectively.  In panel (a), results for a fixed central density $\rho_c=6 \times 10^{14} \text{g}/\text{cm}^3$ ($n_c\simeq2.12n_0$) are shown, while, in panel (b), results for a fixed neutron star mass $M=1.4M_\odot$ are presented. Thick lines represent physical configurations where the angular velocity is below the Kepler (mass-shedding) limit, while dotted lines indicate unphysical configurations where the computed angular velocity exceeds the general relativistic Kepler frequency \eqref{Eq:Keplar_GR}. Results are shown for five Skyrme parameter sets: BSk24 (solid line), SkI4 (dashed line), SkM* (dotted line), SkT5a (dash-dotted line), and SLy4 (two-dot-chain line).
    }
    \label{fig:omega_vs_rratio}
\end{figure}

\begin{table}[t]
\centering
\caption{The calculated central density, the maximum mass and the equatorial radius at the maximum angular frequency (\cf, Fig.~\ref{fig:omega_vs_rratio}) are shown for five EoS parameters. The upper part corresponds to the results for a fixed central density ($n_\text{c}\simeq2.12$), while the lower part corresponds to those for a fixed neutron star mass ($M=1.4M_\odot$).
}\vspace{3mm}
\begin{tabular}{l|cccc}
\hline\hline
\textbf{Model} & $n_c$ [$n_0$]& \boldmath$M_{\text{max}}^{\text{rot}}$ \textbf{[$M_\odot$]} & \boldmath$R_{\text{e}}$ \textbf{[km]} & \boldmath$\Omega_{\text{max}}$ \textbf{[Hz]} \\
\hline
BSk24 & 2.12 & 1.36 & 17.8 & 910 \\
SkI4  & 2.12 & 1.28 & 18.0 & 866 \\
SLy4  & 2.12 & 0.91 & 16.9 & 800 \\
SkT5a & 2.12 & 1.15 & 18.8 & 769 \\
SkM*  & 2.12 & 0.70 & 16.9 & 708 \\
\hline
BSk24 & 1.59 & 1.40 & 18.46 & 871 \\
SkI4  & 1.59 & 1.40 & 18.62 & 851 \\
SLy4  & 1.94 & 1.40 & 17.32 & 920 \\
SkT5a & 1.69 & 1.40 & 18.62 & 807 \\
SkM*  & 2.18 & 1.40 & 16.28 & 960 \\
\hline\hline
\end{tabular}
\label{tab:OmegaMax}
\end{table}

In Fig.~\ref{fig:omega_vs_rratio}(a), we show the angular frequencies obtained for the five EoSs as functions of the axis ratio $R_\text{ratio}$, for a fixed central density of $\rho_c=6\times 10^{14} \text{g}/\text{cm}^3$ ($n_c\simeq2.12n_0$). In the figure, we observe that as the axis ratio decreases, the angular frequency increases gradually, since a smaller axis ratio originates form a stronger centrifugal effect at higher rotational frequencies. Note that there exists a limit to the rotational frequency that can be achieved in equilibrium. Beyond this point, although we could numerically obtain a convergent solution for more oblately deformed configurations, the angular frequency exceeds the general relativistic Kepler (mass-shedding) limit, and are therefore considered as unphysical. Results exceeding this mass-shedding limit are indicated by dotted lines in Figs.~\ref{fig:omega_vs_rratio}(a) and \ref{fig:omega_vs_rratio}(b).

Under the same central density condition, the maximum angular frequency attainable by each EoS parameter set varies. The stiffest EoS, BSk24, allows the highest equilibrium angular frequency within the KEH framework, while the softest EoS, SkM*, yields the lowest (\cf, Table~\ref{tab:nuclear_properties} and Fig.~\ref{fig:EoS}, to see the stiffness). The numerical values of the maximum mass and the equatorial radius are given in Table~\ref{tab:OmegaMax}. As shown in Table~\ref{tab:OmegaMax}, the maximum spin frequencies are achieved at different masses depending on the stiffness: BSk24 supports $1.36M_\odot$, SkI4 $1.28M_\odot$, SLy4 $0.91M_\odot$, SkT5a $1.15M_\odot$, and SkM* only $0.70M_\odot$, at the maximum angular frequency attainable by each EoS. These results confirm that stiffer EoSs can sustain more massive, rapidly rotating stars. Furthermore, although SkT5a has a higher mass than SLy4, it exhibits a lower Kepler frequency due to its larger equatorial radius (18.84\,km \textit{vs.}\ 16.93\,km), demonstrating the radius dependence of the limiting spin. The Kepler (mass-shedding) frequency approximately follows the relation,
\begin{equation}
    \Omega_K \approx \sqrt{\frac{GM}{R^3}},
\end{equation}
which increases with stellar mass and decrease with equatorial radius, indicating its sensitivity to both the mass and equatorial radius of the star.

As mentioned above, the central density and the axis ratio are the inputs of the KEH method and resulting neutron stars have different masses. To better understand the effects of rotation, it is helpful to investigate the structural changes for a specific neutron star with the same mass. In Fig.~\ref{fig:omega_vs_rratio}(b), we show the maximum angular frequencies as functions of the axis ratio under the condition that the gravitational mass is fixed to $1.4M_\odot$. Those values are calculated by repeating structure calculations with the KEH method with varying the central densities to find solutions that correspond to the desired mass. From Fig.~\ref{fig:omega_vs_rratio}(b), we find that the sequence of models capable of sustaining higher rotation rates corresponds exactly to the order of increasing slope parameter $L$: \ie, SkM*, SLy4, BSk24, SkI4, and SkT5a. This trend can be understood as follows: larger values of $L$ generally lead to larger stellar radii, which in turn enhances the effects of rotation. As a result, such stars become more easily deformed and cannot sustain a higher rotation rate before reaching the mass-shedding limit.

\begin{figure}[t]
    \centering
    \includegraphics[width=\linewidth]{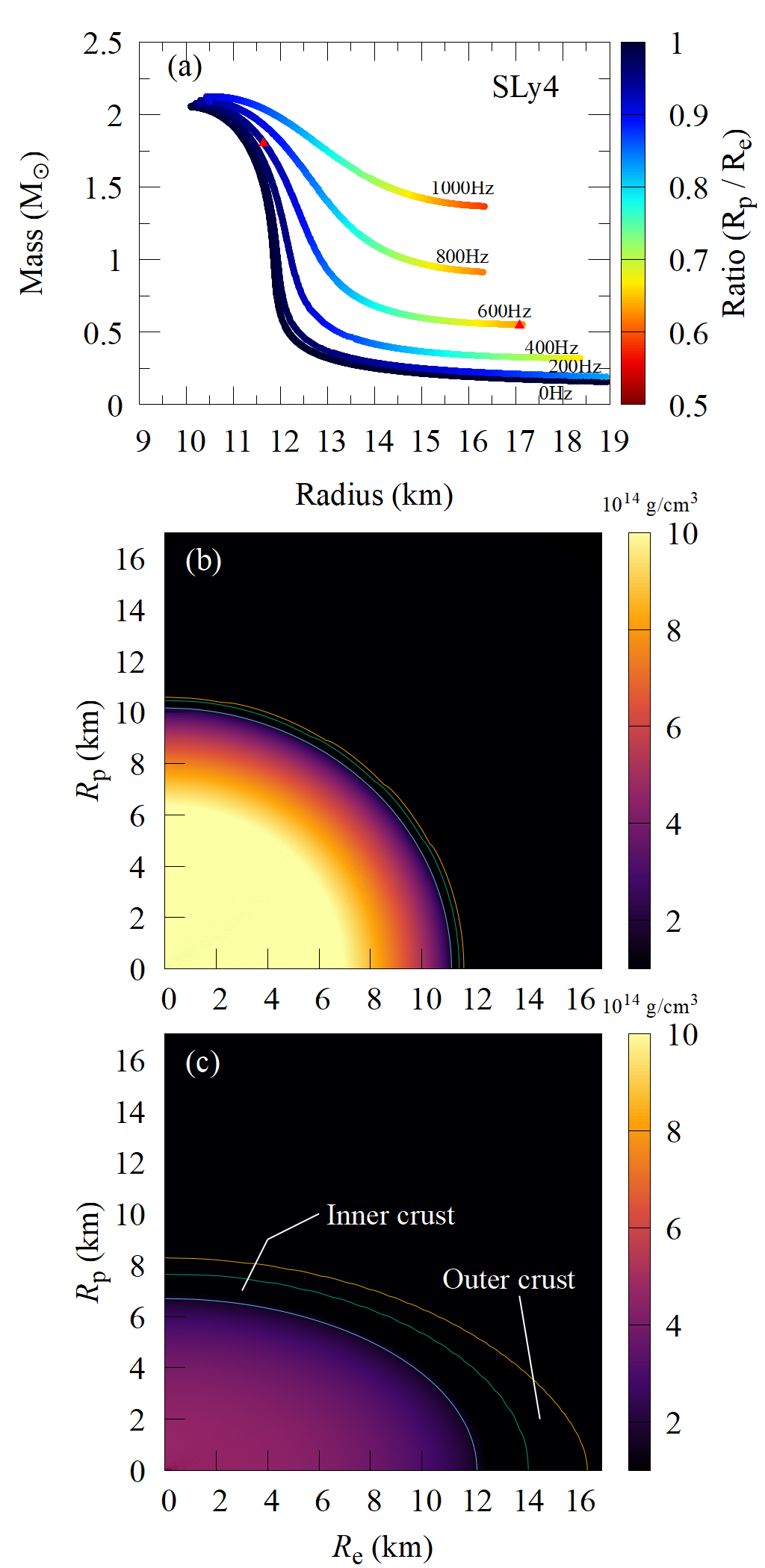}\vspace{-2mm}
    \caption{
    (a) Mass–radius relation of rotating neutron stars at various angular frequency calculated using SLy4 EoS. 
    Red triangles on the $\Omega=600$\,Hz curve indicate two specific cases for which density distributions are shown in (b) and (c), respectively. 
    The three contour lines represent the stellar surface, the transition density between the outer and the inner crust, as well as that between the inner crust and the outer core.
}
    \label{fig:MR_density_combo}
\end{figure}

To more clearly illustrate how rotation affects the internal structure of neutron stars, let us examine stars rotating at various angular frequencies. Figure~\ref{fig:MR_density_combo}(a) shows the mass–radius relations for neutron stars spinning at frequencies from 0\,Hz to 1000\,Hz, calculated using the SLy4 EoS. The non-rotating limit, corresponding to the TOV solution, is represented by the 0\,Hz case. In the rotating case, the equatorial radius $R_\text{e}$ is plotted as its radius. In the figure, the axis ratio $R_\text{ratio}=R_\text{p}/R_\text{e}$ is represented by line colors.

As is evident from the figure, for a given central density, rotation increases both the stellar radius and mass. In other words, the entire $M$–$R$ curve based on the non-rotating TOV solution shifts toward the upper-right direction as rotational frequency increases.

Notably, the changes in structure are small for rotation frequencies below approximately 400\,Hz, but become increasingly significant above this frequency. Higher angular frequencies result in greater deformation. Among neutron stars rotating at the same frequency, those with lower central densities—and thus lower gravitational masses—tend to exhibit stronger rotational deformation. In particular, rapidly rotating neutron stars can reach highly deformed configurations that deviate substantially from their non-rotating counterparts.

To demonstrate this point, we show in Figs.~\ref{fig:MR_density_combo}(b) and \ref{fig:MR_density_combo}(c) density distributions of two selected neutron stars rotating at $\Omega=600$\,Hz, having axis ratios of 0.94 and 0.63, respectively, with different masses and radii, which are indicated by red triangles on the $M$-$R$ curve in Fig.~\ref{fig:MR_density_combo}(a). It is apparent from Figs.~\ref{fig:MR_density_combo}(b) and \ref{fig:MR_density_combo}(c) that a lower-mass star becomes more oblate due to rotational flattening. Intriguingly, neutron stars with lower central densities develop broader low-density regions, resulting in a thicker crust, and the thickness depends on the directions because of the influence of centrifugal forces. That is, the equatorial crust becomes significantly thicker than the polar crust in rotating configurations, as is visible in Fig.~\ref{fig:MR_density_combo}(c). While the crust contributes only a small portion to the global structure in most neutron star models, its relative significance increases in low-mass stars, where it occupies a larger volume fraction. This suggests that the physical properties of the crust may play a more critical role in determining observable features in such rapidly-rotating low-mass stars.

\subsection{Comparison with observational data}

Finally, in this section, we discuss importance of the effects of rotation along with observational data that are often used as astrophysical constraints.

\subsubsection{PSR\,J0740$+$6620}

\begin{figure}[t]
    \centering
    \includegraphics[width=\linewidth]{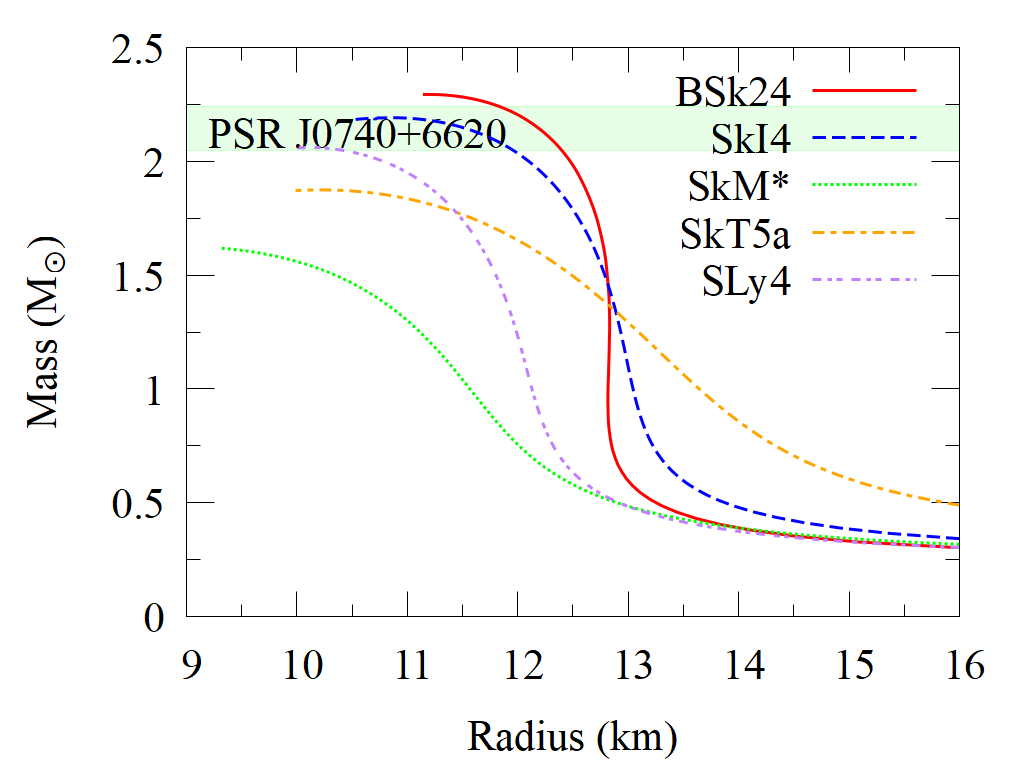}\vspace{-1.5mm}
    \caption{Mass-radius relations for rotating neutron stars with angular frequency of 345\,Hz calculated with five EoS parameters. The green horizontal band indicates the constraint inferred from the observational data of PSR\,J0740$+$6620. Results are shown for five Skyrme parameter sets: BSk24 (solid line), SkI4 (dashed line), SkM* (dotted line), SkT5a (dash-dotted line), and SLy4 (two-dot-chain line).
    }
    \label{fig:MR_346Hz}
\end{figure}

\begin{table}[b]
\centering
\caption{Comparison of neutron star properties from different Skyrme models at 0\,Hz (TOV) and 346\,Hz (KEH). $M^{\text{TOV}}_{\text{max}}$ represents the maximum mass in the TOV calculations, while $M^{\text{KEH}}_{\text{max}}$ represents that in the KEH calculations, and $\Delta M\equiv M^{\text{KEH}}_{\text{max}}-M^{\text{TOV}}_{\text{max}}$. Masses are given in the units of $M_\odot$.}
\begin{tabular}{lccccc}
\toprule
\textbf{Quantity} & \textbf{BSk24} & \textbf{SkI4} & \textbf{SkM*} & \textbf{SkT5a} & \textbf{SLy4} \\
\midrule
$M^{\text{TOV}}_{\text{max}}$ [$M_\odot$] & 2.284 & 2.182 & 1.612 & 1.866 & 2.055 \\
$M^{\text{KEH}}_{\text{max}}$ [$M_\odot$] & 2.295 & 2.192 & 1.618 & 1.875 & 2.062 \\
$\Delta M$ [$M_\odot$]       & 0.012 & 0.010 & 0.006 & 0.0084 & 0.0076 \\
$\Delta M/M_{\text{TOV}}$    & 0.0053 & 0.0046 & 0.0037 & 0.0045 & 0.0037 \\
\bottomrule
\end{tabular}\label{tab:Results_346Hz}
\end{table}

The mass constraint from PSR\,J0740$+$6620 is one of the most stringent observational constraints currently known, indicating a neutron star mass exceeding $2\,M_\odot$. This pulsar is also known to be rotating at a spin frequency of 346\,Hz.
It is to mention that the mass estimate by Cromartie \etal~\cite{Cromartie2020} assumes Schwarzschild masses of the binary system \cite{Damour1986} and works with a post-Newtonian treatment.
In Fig.~\ref{fig:MR_346Hz}, we present the $M$-$R$ curves for rotating neutron stars computed with each EoS parameter set at this angular frequency. Also, maximum masses obtained by the TOV and KEH methods are compared in Table~\ref{tab:Results_346Hz}.

Although the magnitude of rotational effects is small for this frequency, let us discuss it obtained from the KEH calculations. In the case of the SLy4 EoS, the TOV calculation yields a maximum mass of $2.055\,M_\odot$. When rotation effects at 346\,Hz are taken into account by the KEH method, the resulting maximum mass increases to $2.062\,M_\odot$. 
We note that the magnitude of the maximum mass enhancement depends on EoS parameters. For instance, the SkT5a model shows a greater rotational enhancement in the maximum mass than the SLy4 model, despite having a lower non-rotating maximum mass (\cf, Table~\ref{tab:Results_346Hz}). This can be attributed to the significantly larger slope parameter $L$ of SkT5a compared to SLy4. Since $L$ is known to influence the stellar radius, and rotation increases the equatorial radius due to centrifugal effects, the rotational deformation in SkT5a likely leads to a more pronounced increase in maximum mass.

\subsubsection{PSR\,J0030$+$0451}

\begin{figure}[t]
    \centering
    \includegraphics[width=\linewidth]{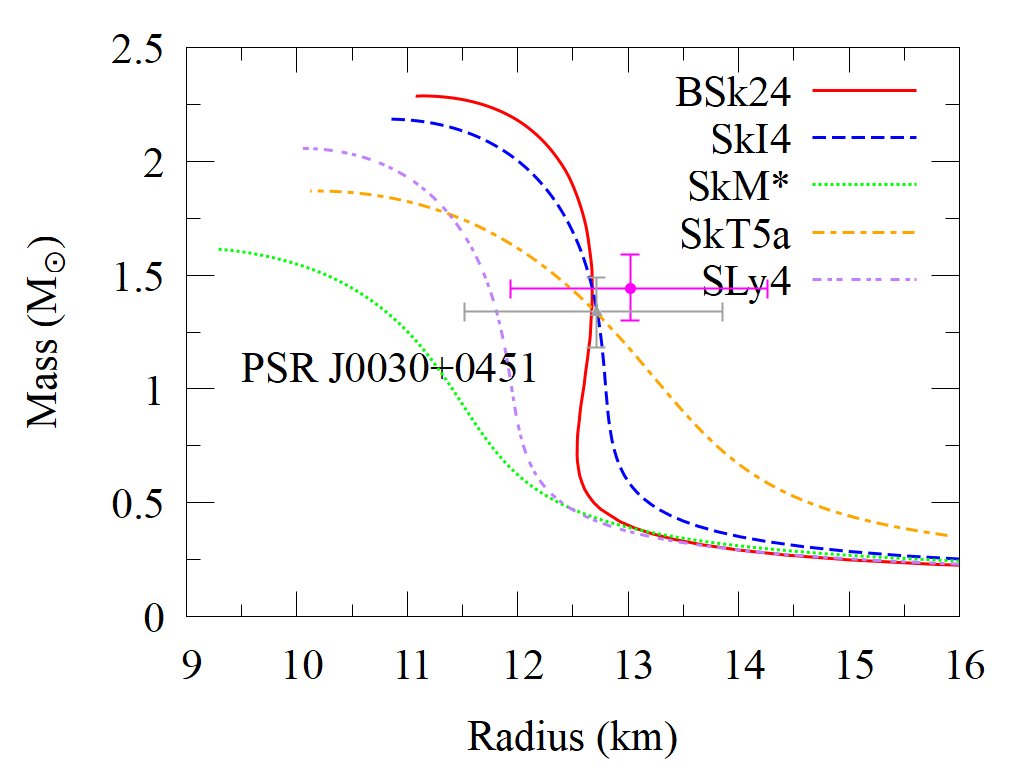}
    \caption{Mass-radius relations for rotating neutron stars with angular frequency of 205\,Hz calculated with five EoS parameters. The gray and pink points with error bars correspond to the constraints inferred from the observational data of PSR\,J0030$+$0451. Results are shown for five Skyrme parameter sets: BSk24 (solid line), SkI4 (dashed line), SkM* (dotted line), SkT5a (dash-dotted line), and SLy4 (two-dot-chain line).
    }\label{fig:MR_205Hz}
\end{figure}

\begin{table}[b]
\centering
\caption{Comparison of neutron star properties from different Skyrme models at 0\,Hz (TOV) and 205\,Hz (KEH). $R_{1.4}^{\text{TOV}}$ represents the radius of a neutron star with $M=1.4M_\odot$ obtained in the TOV calculations, while $R_{1.4}^{\text{KEH}}$ represents that in the KEH calculations, and $\Delta R_{1.4}\equiv R_{1.4}^{\text{KEH}}-R_{1.4}^{\text{TOV}}$. Radii are given in the units of km.}
\begin{tabular}{lccccc}
\hline\hline
\textbf{Quantity} & \textbf{BSk24} & \textbf{SkI4} & \textbf{SkM*} & \textbf{SkT5a} & \textbf{SLy4} \\
\hline
$R^{\text{TOV}}_{\text{1.4}}$ [km] & 12.598 & 12.616 & 10.590 & 12.485 & 11.713 \\
$R^{\text{KEH}}_{\text{1.4}}$ [km] & 12.674 & 12.694 & 10.634 & 12.575 & 11.770 \\
$\Delta R$ [km]       & 0.076 & 0.078 & 0.044 & 0.090 & 0.057 \\
$\Delta R / R_{\text{TOV}}$  & 0.00603 & 0.00618 & 0.00415 & 0.00721 & 0.00487 \\
\hline\hline
\end{tabular}\label{tab:Results_205Hz}
\end{table}

We have also performed calculations at a spin frequency of 205\,Hz, corresponding to PSR\,J0030$+$0451, which is often referenced in connection with radius constraints near the canonical neutron star mass of $1.4\,M_\odot$. Observational analysis of PSR\,J0030$+$0451 is based on two Bayesian studies using data from the NICER mission~\cite{Riley2019,Miller2019}. Although this frequency is relatively low, and thus induces only minor rotational effects, a slight increase in stellar radius was observed in all models. The calculated $M$-$R$ curves for neutron stars with 205\,Hz angular frequency are shown in Fig.~\ref{fig:MR_205Hz}. Also, the equatorial radii for a $1.4\,M_\odot$ neutron star obtained by the TOV and KEH methods are compared in Table~\ref{tab:Results_205Hz}.

Notably, we find that the degree of radius enhancement due to rotation is correlated consistently with the slope parameter \( L \), which is known to influence the neutron star radius. Parameter sets with larger \( L \) values (\cf, Table~\ref{tab:resolutions}) exhibit systematically greater increases in radius under rotation, even at this modest spin frequency.

\subsubsection[PSR J1748--2446ad]{PSR\,J1748--2446ad}

\begin{figure}[t]
    \centering
    \includegraphics[width=\linewidth]{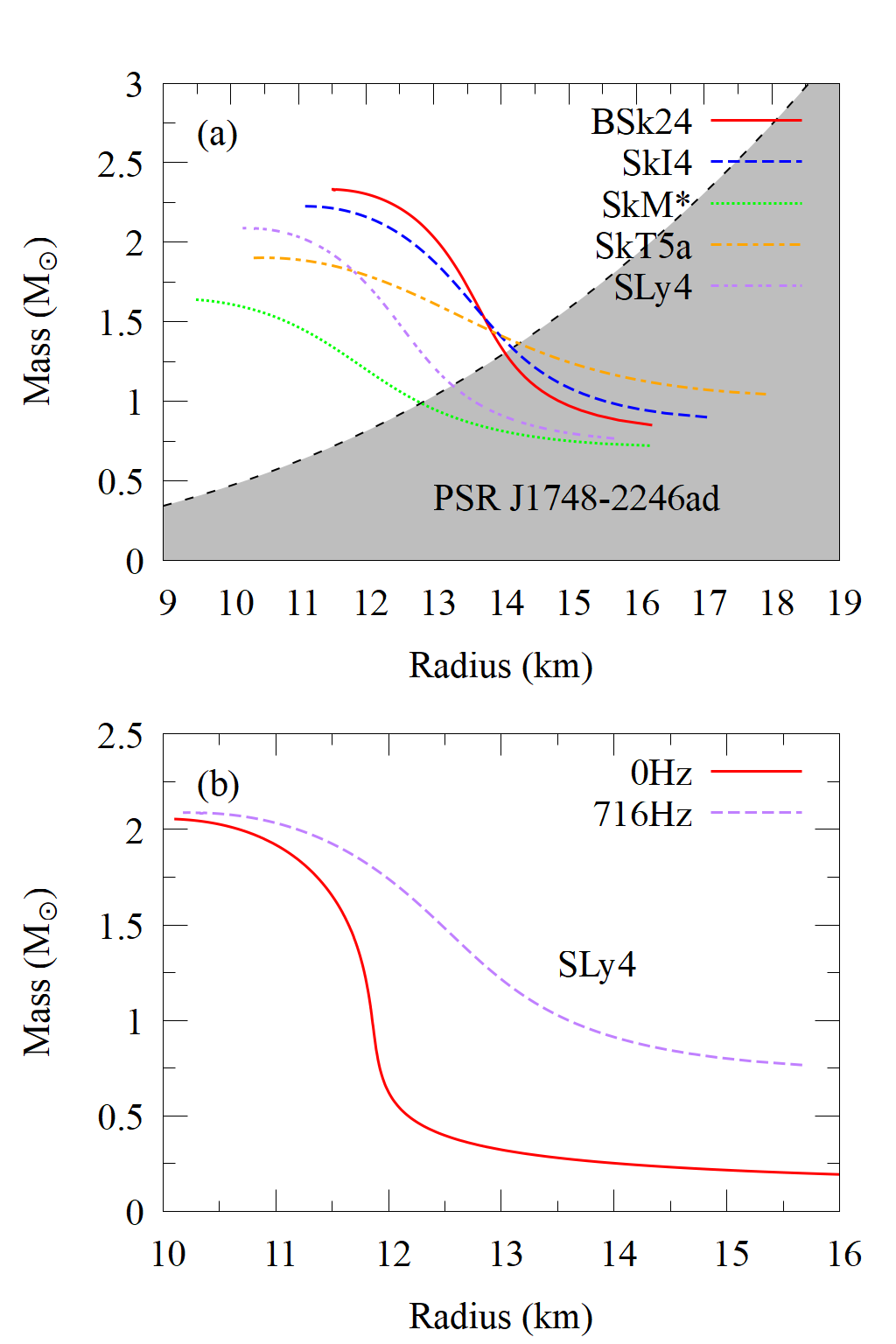}
    \caption{(a) Mass-radius relations for rotating neutron stars with angular frequency of 716\,Hz calculated with five EoS parameters. The gray shaded area indicates the region which is excluded by an empirical upper limit of Eq.~\eqref{Eq:Kepler_empirical}. It is apparent that substantial portion of general relativistic solutions enters inside this region, indicating necessity of modifying the empirical formula. Results are shown for five Skyrme parameter sets: BSk24 (solid line), SkI4 (dashed line), SkM* (dotted line), SkT5a (dash-dotted line), and SLy4 (two-dot-chain line). (b) Comparison between the non-rotating (0\,Hz) and rotating (716\,Hz) cases using the SLy4 EoS, highlighting structural changes induced by rotation.
    }
    \label{fig:MR_716Hz}
\end{figure}

Lastly, we consider the so-called Keplerian constraint derived from PSR\,J1748--2446ad, which has the highest known spin frequency of 716~Hz. Lattimer \etal~\cite{Lattimer2004} suggested an empirical upper limit on the rotation frequency for rigid Newtonian spheres, which is given by
\begin{equation}
\nu_K \approx 1045 \left( \frac{M}{1.4\, M_\odot} \right)^{\!1/2} \left( \frac{10\, \mathrm{km}}{R} \right)^{\!3/2} \ \mathrm{Hz}.
\label{Eq:Kepler_empirical}
\end{equation}
Using this expression, one can draw a constraint curve on the mass–radius diagram, as shown in Fig.~\ref{fig:MR_716Hz}(a). The region below this curve (shaded in gray) is typically regarded as forbidden, according to Eq.~\eqref{Eq:Kepler_empirical}, meaning that neutron stars in this area cannot support stable rotation at 716\,Hz. However, our fully general relativistic calculations using the KEH method demonstrate that a significant portion of this nominally forbidden region can actually correspond to stable rotating neutron star configurations. Our results suggest that the theoretically accessible parameter space for rotating neutron stars is broader than previously estimated. Therefore, more conservative interpretations of the 716\,Hz constraint curve may be necessary. This has important implications for extending the range of observable neutron star configurations in future studies.

The 716\,Hz spin rate is among the highest observed and lies in a regime where relativistic effects become non-negligible. In this frequency range, the differences between the Hartle–Thorne slow-rotation approximation and fully relativistic methods such as KEH are substantial. As shown in Fig.~\ref{fig:MR_716Hz}(b), neutron stars rotating at 716\,Hz exhibit significant changes in their $M$-$R$ relations. Similar behavior was already mentioned at 600\,Hz in Fig.~\ref{fig:MR_density_combo}, where strong rotational deformation leads to markedly increased equatorial radii due to centrifugal flattening at high rotation frequencies. 

Furthermore, in Table~\ref{tab:716Hz}, we compare the neutron star properties at the edge of the constraint curve derived from the TOV solution and those obtained from KEH calculations. We find not only differences in equatorial radius, but also in central density and total mass. These discrepancies emphasize that fully relativistic rotation must be considered when interpreting high-frequency pulsar constraints.

\begin{table}[t]
\centering
\caption{Comparison of neutron star properties from different Skyrme models. 
Upper part shows the minimum central density configurations in rotating stars (KEH), while lower part shows static (non-rotating) stars (TOV) whose mass and radius lie on the 716\,Hz constraint line. Masses, radii, central densities are given in the units of $M_\odot$, km, and $10^{15}$\,g/cm$^3$, respectively.
}\vspace{2mm}
\begin{tabular}{l|cccccc}
\hline\hline
 & \textbf{Quantity} & \textbf{BSk24} & \textbf{SkI4} & \textbf{SkM*} & \textbf{SkT5a} & \textbf{SLy4} \\
\hline
\multirow{3}{*}{\makecell[l]{\textbf{KEH}\\(min $\rho_c$)}} 
 & $n_c$ & 0.47 & 0.49 & 0.61 & 0.54 & 0.54 \\
 & $M$      & 0.83 & 0.89 & 0.72 & 1.04 & 0.75 \\
 & $R$      & 16.9 & 17.4 & 16.4 & 18.3 & 16.5 \\
\hline
\multirow{3}{*}{\makecell[l]{\textbf{TOV}\\(at 716\,Hz threshold)}} 
 & $n_c$  & 0.63 & 0.67 & 0.83 & 0.81 & 0.72 \\
 & $M$       & 1.15 & 1.18 & 0.88 & 1.23 & 0.97 \\
 & $R$       & 12.5 & 12.7 & 11.5 & 12.8 & 11.9 \\
\hline\hline
\end{tabular}\label{tab:716Hz}
\end{table}

\section{Summary and prospect}\label{Sec:Conclusion}

Currently, the majority of studies of the nuclear equations of state (EoSs) through a relation between neutron star masses and radii relays on the Tolman-Oppenheimer-Volkoff (TOV) equation which describes spherically symmetric, static neutron stars, neglecting effects of rotation. While the rotational effects on neutron star structure are actually small for slowly rotating systems with angular frequencies less than about 400\,Hz, they can not be neglected for rapidly rotating ones, like millisecond pulsars. To describe the structure of rapidly rotating neutron stars, one needs to employ fully relativistic methods that incorporate with centrifugal effects. Aiming at elucidating the effects of rotation on neutron star structures, refining modeling of nuclear matter as well as neutron star structure, we have newly developed our own computational code that calculates structure of rapidly rotating neutron stars in a fully general relativistic manner based on the method proposed by Komatsu, Eriguchi, and Hachisu (the KEH method).

In this study, we constructed EoSs for the $npe\mu$ composition using Skyrme-type nuclear interactions, and discussed how the resulting macroscopic properties of neutron stars relate to the nuclear matter properties near saturation density. Employing the fully general relativistic KEH method, we have investigated the rotational deformation of neutron stars and analyzed how their internal structure and mass–radius ($M$–$R$) relations change with increasing spin. We have found that, for a fixed rotational frequency, stars with lower central densities undergo greater deformation, and for a fixed central density, more rapid rotation leads to stronger flattening. Furthermore, we have identified a clear relationship between the maximum allowable rotational frequency and the stiffness of the EoS.

We have also performed calculations for three commonly referenced pulsar constraints---PSR\,J0740$+$6620, PSR\,J0030$+$0451, and PSR\,J1748--2446ad---at their respective observed spin frequencies. While the first two objects are millisecond pulsars yet with moderately slow rotation, the corrections due to rotation are small but finite, indicating the necessity of incorporating rotational effects when aiming for precise radius constraints. For PSR\,J1748--2446ad, the fastest known pulsar at 716\,Hz, we have compared the empirical Kepler limit with our KEH-based results and showed that a substantial portion of the parameter space traditionally considered forbidden may in fact correspond to stable stellar configurations. This finding suggests the need for more conservative rotational constraints that take full relativistic effects into account.

Although the KEH method provides a robust framework for modeling rotating neutron stars, it shows small dependence on numerical parameters such as the mesh resolution, initial density profiles, and interpolation scheme. In our analysis with five EoS models, successful convergence was achieved through iterative, empirical tuning. However, in order to systematically explore the 716\,Hz constraint across a wider range of EoSs—--especially to identify the lowest-mass neutron star that can stably rotate at this frequency—--it will be necessary to develop a more numerically stable and automated framework, possibly incorporating parallel computation.

Despite the limited number of EoS models examined in this work, we were able to infer some correlations between nuclear matter properties at saturation (\eg, the symmetry energy slope $L$ and rotational deformation. To generalize this relationship, future studies could fix all nuclear matter parameters except one and investigate how variations in that parameter affect the rotational response of neutron stars. This strategy may allow for a more systematic understanding of which nuclear matter properties most strongly influence the shape and structure of rapidly-rotating neutron stars.

In addition, in the recent work by G\"artlein \textit{et al.}~\cite{Gartlein2025}, it was shown, applying the \texttt{RNS} code for hybrid stars, that for the extreme case of the fastest pulsar PSR J1748–2446ad, the deconfinement phase transition occurs at a lower central density compared to the non-rotating case. Therefore, extending the present study to incorporate such internal phase dynamics under rapid rotation would be of clear value.

Finally, this study adopted the simplest case of rigid rotation within the KEH method, which is similar in formulation to the publicly available \texttt{RNS} code. However, it is known that differential rotation can support even larger maximum masses. In such cases, stability analysis becomes crucial, particularly in connection with phenomena such as r-mode instabilities and the viscous properties of dense nuclear matter. These aspects represent promising directions for future research.

\section*{Acknowledgments}
We would like to thank Kenta Yoshimura (Science Tokyo) for careful reading of the manuscript and for providing useful comments on this article. H.K. would like to thank Shunsuke Yasunaga (Science Tokyo) for mathematical support in deriving the integral forms of Poisson-like equations. We are also thankful to Dr.~Jinho Kim (Korea Astronomy and Space Science Institute) and Dr.~Tsuyoshi Miyatsu (Soongsil University) for valuable instructions that were crucial for developing the new computational code. This work is supported by JST SPRING, Japan Grant Number JPMJSP2180, JSPS Grant-in-Aid for Scientific Research, Grants No.~23K03410, No.~23K25864, and No.~JP25H01269. This work was also supported by the Sasakawa Scientific Research Grant from the Japan Science Society, Hattori International Scholarship Foundation.

\bibliography{reference}

\end{document}